# WAGONN: <u>W</u>eight Bit <u>Ag</u>gl<u>o</u>meration in Crossbar Arrays for Reduced Impact of Interconnect Resistance on D<u>NN</u> Inference Accuracy

Jeffry Victor, Dong Eun Kim, Chunguang Wang, Kaushik Roy and Sumeet Gupta
Purdue University, West Lafayette, IN, USA

## ABSTRACT

Deep neural network (DNN) accelerators employing crossbar arrays capable of in-memory computing (IMC) are highly promising for neural computing platforms. However, in deeply scaled technologies, interconnect resistance severely impairs IMC robustness, leading to a drop in the system accuracy. To address this problem, we propose WAGONN - a technique based on agglomerating weight bits in crossbar arrays which alleviates the detrimental effect of wire resistance. For 8T-SRAM-based 128x128 crossbar arrays in 7nm technology, WAGONN enhances the accuracy from 47.78% to 83.5% for ResNet-20/CIFAR-10. We also show that WAGONN can be used synergistically with Partial-Word-Line-Activation, further boosting the accuracy. Further, we evaluate the implications of WAGONN for compact ferroelectric transistor-based crossbar arrays and show accuracy enhancement. WAGONN incurs minimal hardware overhead, with less than a 1% increase in energy consumption. Additionally, the latency and area overheads of WAGONN are ~1% and ~16%, respectively when 1 ADC is utilized per crossbar array.

This work was supported in part by the Center for Brain-Inspired Computing (C-BRIC) and in part by the Center for the Co-Design if Cognitive Systems (COCOSYS), funded by the Semiconductor Research Corporation (SRC) and DARPA under Grant AWD-004311-S4. (*Corresponding author: Jeffry Victor*.)

Jeffry Victor, Dong Eun Kim, Chunguang Wang, Kaushik Roy and Sumeet Gupta are with the Elmore School of Electrical and Computer Engineering, Purdue University, West Lafayette, Indiana, USA - 47906. (e-mail: louis8@purdue.edu, kim2976@purdue.edu, wang4015@purdue.edu, kaushik@purdue.edu, guptask@purdue.edu)

## 1 INTRODUCTION

The remarkable success of deep neural networks (DNNs) in achieving super-human accuracy for several tasks has led to a surge in the exploration of DNN accelerators [1], [2]. However, the current design approaches for DNN hardware involving von-Neumann platforms (such as central/graphic/tensor processing units - CPUs/GPUs/TPUs) are plagued by power-hungry and performance-limiting memory-processor transactions, which make them ill-suited for data-intensive DNN workloads [3].

To address these limitations, in-memory computing (IMC) has emerged as a promising technique to integrate the storage and processing of data in a memory macro, alleviating the von-Neumann bottleneck [1], [2], [4], [5]. A common IMC approach employs a crossbar memory array to compute matrix-vector multiplication (MVM) of activations and weights [1]. This approach reduces the need to transfer large weight and activation matrices between memory and logic, leading to substantial improvements in performance compared to CPUs, GPUs and TPUs [6], [7].

Despite the immense promise of IMC, the challenges posed by the presence of non-idealities thwarts its adoption in mainstream applications. In an ideal scenario, the conductance of the memory element (storing the weight $w$) and the word-line voltage (representing the input activation $In$) produce a current which encodes the scalar product of $In$ and $w$. This current is naturally summed up on the sense-line (SL) to produce the MVM output. However, hardware non-idealities such as wire resistance, driver/sink resistance and device non-linearities/variations lead to a deviation of the SL current from its expected value. If large, this deviation can lead to computational errors, impairing DNN accuracy [1], [8].

This issue becomes even more critical in deeply scaled technologies. The importance of scalability of DNN accelerators is paramount to support the ever-increasing size of the DNN models for handling complex tasks. The technological advancements have led to scalable transistor topologies and non-volatile memory devices such as ferroelectric transistors (FeFETs) that are amenable to energy efficient IMC. However, interconnect scaling has been a major challenge [9], [10], [11].

Conventional copper (Cu) interconnects require barrier/liner layers to mitigate the issues associated with electromigration [10]. However, these layers do not scale well, thereby reducing the percentage of Cu as technology is scaled. This not only reduces the active area for current conduction, but also leads to an increase in sidewall scattering, increasing the interconnect resistivity (*not just resistance*) [3]. This issue is particularly severe for IMC since the wires in the crossbar array need to carry the accumulated currents of multiple memory cells, leading to large IR drops and significant computational errors.

Recognizing the significance of this issue, several techniques are being explored at the technology [10], [11] as well as design levels [12], [13], [14], [15], [16], [17]. For the former, alternate materials and processes for interconnects are being investigated, each with its own pros and cons [10], [11]. To complement these, several circuit design and algorithmic innovations promise to further enhance IMC robustness. These include hardware-aware training [12], [13], [14], [15], re-distributing the weights [17] and partial word-line activation (PWA) [16].

While promising, these solutions have their own costs (details later). Further, most of these approaches have been analyzed for technologies in which the wire resistance is small, and the associated non-idealities are fairly manageable. While the scalability of these techniques needs further analysis, at the same time, there is a need for new approaches that effectively mitigate the detrimental effect of the wire resistance on IMC robustness and DNN accuracy in deeply scaled technologies.

In this work, we address this need by proposing a new technique based on <u>w</u>eight bit <u>ag</u>glomeration in crossbar arrays for enhancing D<u>NN</u> accuracy, or WAGONN. Our approach re-maps weight bits in the crossbar arrays to mitigate the effect of wire resistances on IMC robustness. WAGONN offers a training-free approach with minimal hardware overheads. Our key contributions are as follows:

-We propose WAGONN which re-maps the weights in the crossbar arrays in such a way that the IR drops in the wire resistance is significantly reduced, thereby achieving high DNN accuracy in the presence of hardware non-idealities.

-We establish the efficacy of WAGONN in deeply scaled technologies (7nm node) by analyzing 8T-SRAM and FeFET-based crossbar arrays.

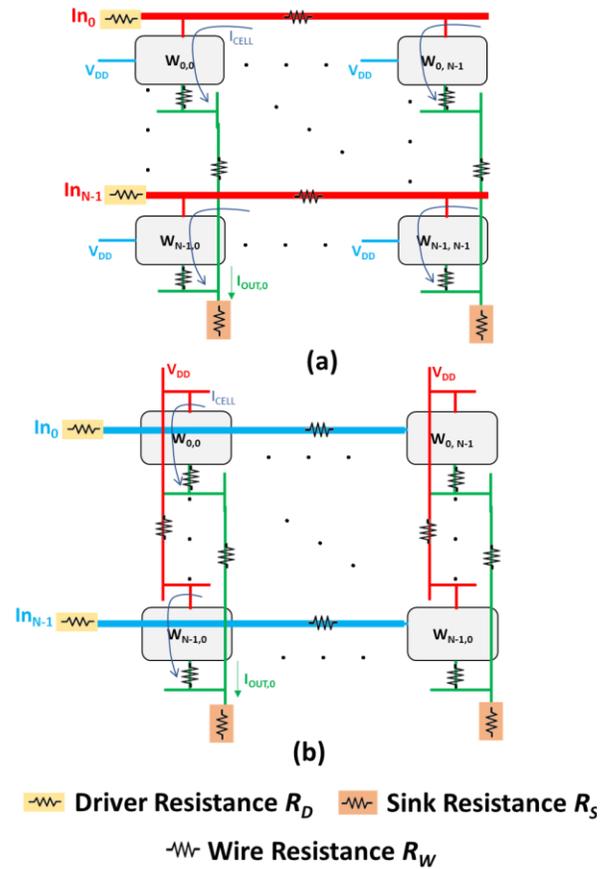

Fig. 1 Input configurations (a) D-input and (b) G-input

-We show the synergy between WAGONN with PWA [16] (used to manage analog-to-digital converter (ADC) costs in IMC macros).

-We establish the resilience of WAGONN to device-to-device variations by analyzing its performance for varying levels of variations.

-We analyze the hardware implications of WAGONN which requires dynamic activation re-mapping along with weight re-mapping to maintain the MVM functionality. We show that the design overheads of WAGONN are minimal.

## 2 RELATED WORK

Several works [7-12] have explored solutions to mitigate the effect of hardware non-idealities that stem from parasitic resistances, device non-linearities and process variations. The works in [12], [13] have shown the efficacy of training DNNs including the device variations and wire resistance to improve their inference accuracy. However, these studies primarily focus on small datasets and networks. The authors in [14] mitigated hardware non-idealities by re-training the network while limiting the output current range. While promising, these works [6-8] incur the expensive cost of re-training the DNN.

Several other works have explored training-free strategies to counter non-idealities. In [17], crossbar columns are rearranged to position high current-producing or accuracy-critical columns closer to the drivers, thereby reducing non-idealities. Another work in [16] utilizes partial wordline activation (PWA) to improve the inference accuracy by reducing the SL current. PWA asserts only a subset of WLs in one cycle, trading off inference accuracy with parallelism. In addition to mitigating non-idealities, it reduces the overheads of ADCs.

Another work in [15], combines the approach of training the network with the approach of rearranging columns to recover accuracy lost due to pruning.

The works discussed above span technology nodes from 45nm to 65 nm, with wire resistance ranging from 2 Ω to 10 Ω per bitcell. However, in deeply-scaled nodes such as 7nm, the wire resistance can increase up to 20 Ω per bitcell [8], [9], significantly impairing DNN accuracy. Hence, the efficacy of the techniques proposed in [7-12] need further evaluation for deeply-scaled technologies.

Recently, Wang et.al in [8] examined non-idealities in highly-scaled (7nm) technology nodes for DNN inference. Their work showed that G-input design i.e. applying the inputs on the gate of FeFETs or the access transistors of SRAMs reduces the impact of non-idealities compared to the D-input design i.e. applying the inputs on the drain terminals (Fig. 1). In D-input (Fig. 1a), Bit Line (BL) is connected horizontally while the source is connected vertically along the SL. Thus, the current in each bit-cell is affected by IR-drops along both these lines. Since the BL and SL are orthogonal, the current through any bit-cell is directly dependent on the conductance of bit-cells in its row and column, and, indirectly, on all the bit-cells in the array (through sneak current paths). On the other hand, for G-input (Fig. 1b), current-carrying BL and SL are routed vertically along the column, and hence, the current in a bit-cell is dependent on the bit-cells only in that column. This reduction of data dependency for the G-input lowers the range of current for each output and reduces the impact of the parasitic resistances.

For a further improvement in DNN accuracy in deeply-scaled technology nodes, we propose a training-free approach called WAGONN which targets the mitigation of interconnect-induced non-idealities in crossbar arrays. Our technique complements some of the aforementioned

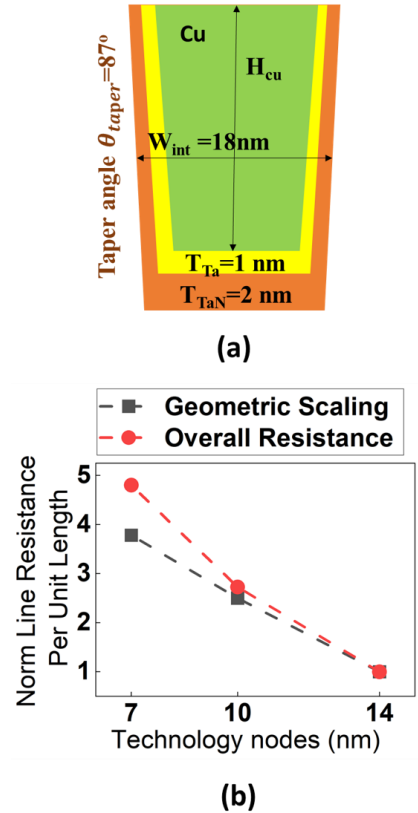

Fig. 2 (a) Interconnect Structure (b) Normalized Line Resistance per unit length across nodes

solutions such as G-input design and PWA (discussed later) and may be used in conjunction with other prior approaches.

## 3 CROSSBAR ARRAY ANALYSIS AT 7 nm TECHNOLOGY NODE

Before we present WAGONN, we analyze IMC in crossbar arrays at the 7nm technology node in this section to lay the groundwork for the rest of the paper. We discuss the increasing impact of interconnect resistance with technology scaling and illustrate its impact on IMC robustness.

### 3.1 Interconnect Modeling and Analysis

Fig. 2 (a) shows the cross-section of the interconnect from [9] that we utilize for our analysis in this work. The active current conducting portion is composed of copper (Cu), which is surrounded by Ta liner and TaN barrier. In state-of-the-art technologies, 2nm Ta liner and 2nm TaN barrier is utilized [9]. However, scaling the liner to 1nm is being explored to increase the cross-sectional area of Cu for higher conductance [9].

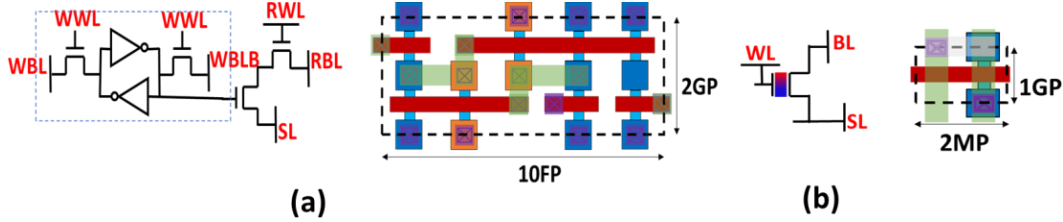

Fig. 3 Schematic and layout of (a) 8T-SRAM (b) FeFET

In this work, we utilize the scaled liner design with 1nm Ta and 2nm TaN. We assume 50% sidewall coverage and a sidewall taper angle of $87^0$, based on [9]. It may be noted that for the state-of-the-art interconnects (with 2nm liner), the interconnect-related issues discussed later will be aggravated and the benefits of the proposed WAGONN technique is expected to be even more significant.

To model the interconnects, we include the sidewall and grain-boundary (GB) scattering based on [18] and [19], respectively. For the 7nm technology node, we use the interconnect width (corresponding to M1-M3) to be 18nm and the pitch of 36nm. We build 3D interconnect models in COMSOL incorporating sidewall and GB scattering (parameters summarized in Table I) to analyze the structures in Fig. 1 (a).

Based on our models, we observe that the overall line resistance increases by 4.8x as we scale the technology from the 14 nm to 7 nm (Fig. 1b). Notably, geometric scaling accounts only for a 4.0x increase in wire resistance. The additional increase is due to increased sidewall scattering, which arises from the reduced copper cross-sectional area as we scale from 14 nm to 7 nm [4], [14]. This indicates that future technology scaling beyond 7 nm will further exacerbate wire resistance due to aggravated scattering, making accuracy-enhancing techniques, such as WAGONN, even more pertinent for maintaining performance.

From our models, we obtain the values for the line resistance to be 182 Ω/μm and via resistance to be 78 Ω. These values are in good agreement with prior works [9], [20]. The line resistance also matches well with an IBM technology [10].

### 3.2 Simulation of Crossbar Non-Idealities and DNN Accuracy

In this section, we analyze crossbar arrays utilizing 8T-SRAM and FeFET designs [2] (Fig. 3). For 8T SRAMs, we utilize the predictive technology models (PTM) corresponding to 7 nm low standby power (LSTP) FinFETs [21]. For FeFETs, we utilize the model in [22] based on Miller's multi-domain equations for the ferroelectrics (emulating the Preisach model). We couple the Miller's model with 7nm PTM models to obtain the polarization-dependent device characteristics [23]. The parameters for the devices are summarized in Table I.

We design the 8T SRAM and FeFET bit-cells considering G-input configuration and draw their layouts (Fig. 3) based on the metal-pitch and gate-pitch values for 7nm Intel process [24]. We consider a bit-slice of 1 in each bit-cell and input bit-stream = 1. The conductance values for various input ($In$) and weight bit ($w$) combinations (obtained from our models) are provided in Fig. 4 (a).

We observe that compared to FeFETs, SRAMs offer higher ratio of conductance for the ON ($In$=1 and $w$=1) and OFF ($In$=0 or $w$=0) bit-cells, while FeFETs offer a more compact bit-cell footprint (half the cell height). As we will

|  | SRAM (S) | FeFET (S) |
|---|---|---|
| In = 1 W = 1 | $1.6 \times 10^{-5}$ | $1.6 \times 10^{-5}$ |
| In = 1 W = 0 | $4.7 \times 10^{-12}$ | $2.50 \times 10^{-7}$ |
| In = 0 W = 1 | $6.6 \times 10^{-12}$ | $4.3 \times 10^{-8}$ |
| In = 0 W = 0 | $2.2 \times 10^{-12}$ | $2.0 \times 10^{-10}$ |

(a)

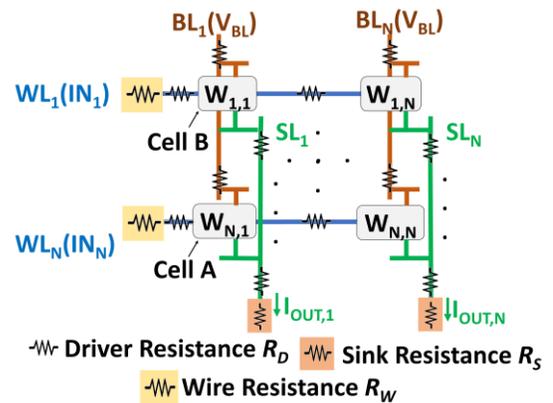

(b)

Fig. 4 (a) Bit-cell conductance for different input-weight combinations (b) Crossbar with its non-idealities discuss later, both these factors play an important role in IMC. Based on these designs, we utilize the GENIEX framework [1] to study the effect of crossbar non-idealities on DNN inference accuracy at the 7nm technology node.

### 3.3 Non-idealities in Crossbar Arrays

Fig. 4 (b) shows a typical crossbar array along with the parasitic resistances (wire resistance $R_W$, driver resistance $R_D$ and sink resistance $R_S$). As the current-carrying lines (RBL/BL and SL) are routed along the columns, their length and therefore, $R_W$ is determined by the layout height of the cells and the number of cells in a column. The IR drop in the parasitic resistances ($R_W$, $R_S$ and $R_D$) lead to deviation of the output current from its expected values, which can result in computational errors. Due to high $R_W$ at the 7nm technology node (discussed in the previous sub-section), these deviations are large, leading to a large impact on DNN accuracy.

Our analysis shows that the impact of $R_W$ is twofold. First, the drain voltage of the transistors is reduced. Second, the source voltage is increased (leading to source degeneration). While both contribute in reducing the cell current, the latter effect is more critical in SRAMs and FeFETs as the increase in source voltage reduces both the gate-to-source voltage ($V_{GS}$) and drain-to-source voltage ($V_{DS}$).

Moreover, due to the distributed nature of $R_W$ in a column, the deviation in the output current depends on the distribution of the weights and inputs in a column. For instance, consider two cells which are ON (i.e. with $In$=1 and $w$=1): Cell A is closer to the ADC (bottom of the array in Fig. 4 (a)) and farther away from the BL/RBL driver than Cell B. During IMC, the source voltage of Cell A is closer to 0 (the ideal value), but its drain voltage is lower than Cell B. On the other hand. Cell B suffers from stronger source degeneration. As mentioned before, the effect of source degeneration is more dominant, and hence, the deviation of current in Cell A is lower than Cell B. The standard weight mapping leads to different input/weight distributions in crossbar arrays, resulting in range of output currents for the same expected MVM product.

Our simulations for 128x128 (64x64) arrays at 7nm technology node show that high $R_W$ (along with other hardware non-idealities) causes the DNN accuracy to drop from the software accuracy of 92.8% to 47.78% (88.8%) for 8T-SRAMs and 84.85% (91.63%) for FeFETs in a. Note, SRAMs show a higher accuracy drop than FeFETs as their cell height, and therefore, $R_W$ is larger.

**TABLE I**
Device & Interconnect Parameters

| | |
|---|---|
| $\rho_{bulk}[\Omega \cdot \mu m]$ | 0.0172 |
| $\rho_{Ta}[\Omega \cdot \mu m]$ | 2 |
| $\rho_{TaN}[\Omega \cdot \mu m]$ | 3 |
| Grain-boundary reflection coefficient | 0.135 |
| Electron mean free path [nm] | 40 |
| FE Thickness [nm] | 7 |
| Saturated Polarization [$\mu C/cm^2$] | 30 |
| Remanent Polarization [$\mu C/cm^2$] | 27 |
| FE Relative Permittivity | 22 |
| Coercive Electric Field [MV/cm] | 2.4 |
| Fin Pitch (FP) [nm] | 27 |
| Gate Pitch (GP) [nm] | 54 |
| Metal Pitch (MP) [nm] | 36 |
| WL Voltage [V] | 0.7 |
| BL Voltage [V] | 0.25 |

## 4 WAGONN: Agglomerating weights to reduce wire-induced non-idealities

### 4.1 The key idea

The discussions in the previous section suggest that due to IR drops on the parasitic resistances and the consequent source degeneration, an 8T SRAM or FeFET bit-cell farther away from the ADC exhibits a larger reduction in its current (encoding the scalar product of its weight and input) than the one closer to the ADC. Hence, we propose to reduce the deviation of the dot product (SL current) from its expected value by re-mapping the weights in the array such that the bit-cells storing $w$ =1 (low resistance state or LRS) are agglomerated in the lower rows (closer to the ADC), while the bit-cells with $w$=0 (high resistance state or HRS) are in the upper rows. Interestingly, the proposed technique not only increases the current of the ON bit-cells ($w$=1 and $In$ =1) towards the expected values (by lowering the effect of IR drops), it also reduces the current in the bit-cells storing 0 by inducing a stronger source degeneration. Note that bit-cells with $w$=0 produce non-zero currents due to finite resistance of the HRS. This aggravates the non-ideal effects in crossbar arrays. WAGONN reduces the current in such bit-cells, thereby further mitigating the non-idealities. It is also noteworthy that when the input ($In$) = 0, the LRS bit-cells also produce non-zero current due to leakage. However, these leakage currents are much smaller than the currents corresponding to HRS of the bit-cell. Hence, reducing the current for $w$=0 via the proposed technique has a more dominant and desirable effect compared to the leakage increase in the cells storing 1. Thus, WAGONN achieves an increase in the ON current and the reduction in the HRS currents at an insignificant cost of leakage increase.

To maintain the correct MVM functionality, the proposed weight re-mapping must be accompanied by re-mapping of the input activations. Since the inputs are applied along the rows of a crossbar array, the columns cannot be re-arranged independently. Therefore, we propose to re-arrange the weights by analyzing the sum of weights stored in a row (called the row-sum) and mapping the rows with the largest row-sum to the bottom of the crossbar array. In this way, the columns of the crossbar array are expected to have a large number of LRS bit-cells closer to the ADC.

To achieve this, we follow the following process (Fig. 4). First, we compute the row-sum of each row. Second, the row-sum of each row is concatenated into a vector and sorted in ascending order. In this step, we use an additional vector called the tracking vector which stores the position of each row in the sorted row-sum vector. For example, if the second row has the highest row-sum out of all 64 rows, the tracking vector updates its second entry as 64. Essentially, the tracking vector stores the address of the destination that each row should be swapped into while deploying the weights in the crossbar array. These two steps are performed in software *before* mapping the weights in the DNN accelerator incurring a one-time cost. Next, the re-arranged weights are deployed in the crossbar arrays. The corresponding inputs are dynamically swapped according to the tracking vector in the hardware. We will discuss the hardware implications of this later. After this step, the crossbar arrays have most of the LRS bit-cells in the bottom rows.

In this work, we re-map the rows for each crossbar array independently. Alternatively, we could perform this swap at a higher granularity by taking row-sums across multiple crossbars and then performing weight re-mapping based on a common tracking vector. This would reduce the hardware overhead and is expected to simplify dynamic re-mapping of the inputs. However, this would, in general, reduce the effectiveness of WAGONN in mitigating the effect of non-idealities (which needs to be analyzed in the future). In this work, we focus on independent re-mapping for each crossbar array, showing its effectiveness in enhancing DNN accuracy and analyzing its hardware costs.

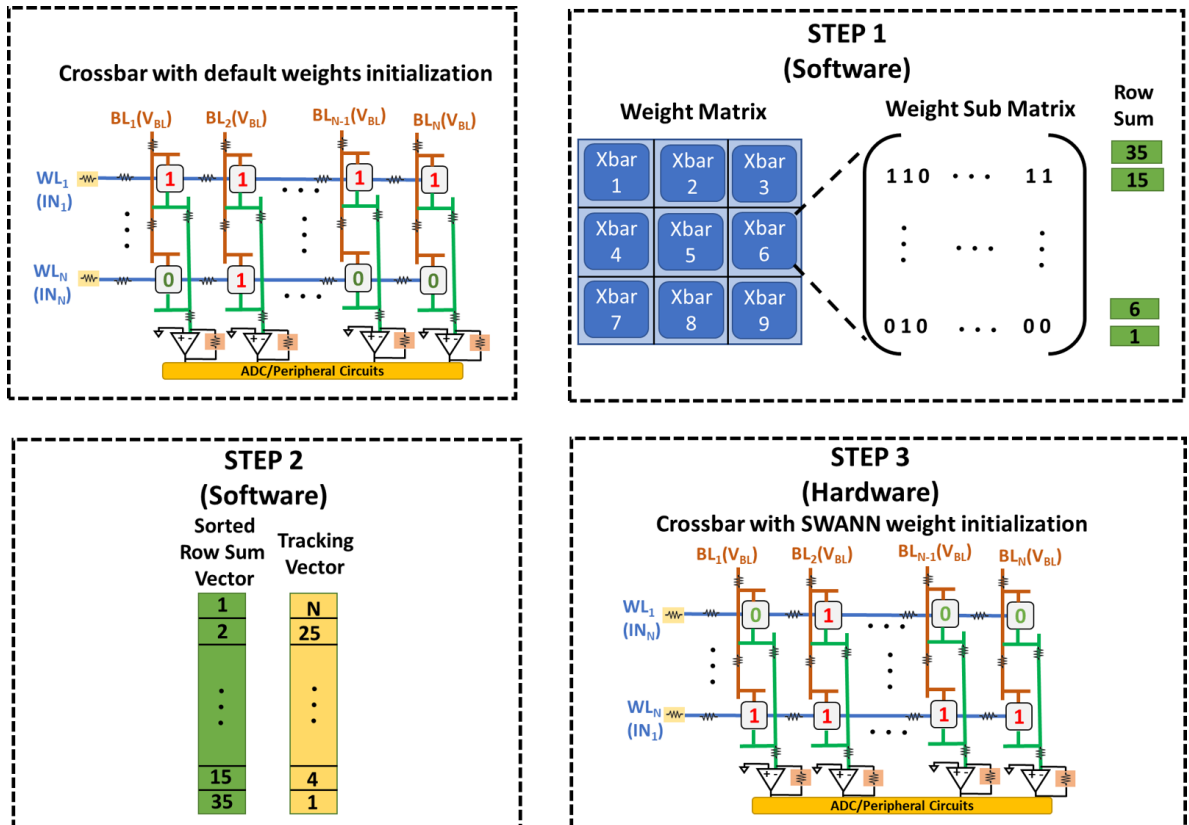

Fig. 5 Three Steps of the WAGONN technique

## 4.2 Complementing Partial Wordline Activation (PWA) with WAGONN

PWA is a common technique used for IMC in crossbar arrays, in which only a subset of rows is activated in a single cycle. Thus, the MVM computation is broken down in multiple steps. This helps in mitigating two important issues. First, the range of the SL current is reduced, mitigating the IR drops and the associated non-idealities [16]. Second, this relaxes the bit-precision requirements for the ADCs, significantly reducing their energy and area costs (which are known to be dominant in IMC macros [25], [26]).

In standard methods (where the weight and input distribution is arbitrary), each cycle of PWA commonly involves asserting consecutive rows concurrently. For example, for a crossbar of size $NxN$ grouped into $M$ groups the first $N/M$ rows are asserted in the first cycle, second $N/M$ rows in the second cycle and so on. Interestingly, in WAGONN, the distribution of cells storing weight bit=1 is concentrated towards the bottom of the array due to weight agglomeration. Thus, if standard PWA is naively applied with WAGONN, the current in the later cycle is expected to be much more, which could reduce the effectiveness of PWA.

Interestingly, since the weight pattern is less arbitrary in WAGONN, we can synergize the beneficial effects of PWA and WAGONN. For this, instead of asserting consecutive rows in a single cycle, we propose activating the rows in a distributed fashion in a WAGONN crossbar. For example, in the crossbar with $N$ rows and $M$ groups. The $i$th group asserted in the $i$th cycle consists of rows $i$, $i+N/M$, $i+2N/M$, etc. We will refer to this approach as Distributed PWA (DPWA). By grouping WLs in such a distributed fashion, we distribute the high row-sum rows among different groups. Thus, in each DPWA cycle, the number of cells producing scalar output of 1 is expected to reduce compared to a naïve PWA+WAGONN, thereby further mitigating the effect of non-idealities.

## 5 RESULTS AND DISCUSSION

In this section, we quantify the efficacy of WAGONN in boosting the inference accuracy and analyze the associated hardware overheads. We analyze ResNet-20 for CIFAR-10 dataset and compare the proposed WAGONN technique with the baseline (in which standard weight mapping is used).

### 5.1 Effect of WAGONN on Accuracy

The effect of WAGONN on accelerators using 64x64 and 128x128 8T SRAM arrays is shown in Fig. 5 (a). For the baseline design, 64x64 SRAM crossbar arrays yield an inference accuracy quite close to the software baseline. However, as we increase array size to 128x128, the inference accuracy drops by 41% due to higher wire resistance. On applying the proposed WAGONN technique, we observe the inference accuracy increases (by 2.85%) from 88.8% to 91.65% in the 64x64 array while it increases (by 35.72%) from 47.78% to 83.5% in the 128x128 array. In other words, by reducing the impact of interconnect-induced non-idealities, WAGONN can enable the design of *larger* crossbar arrays in deeply scaled technologies to lower the peripheral circuit overhead.

To gain further insights into WAGONN, we show its implications for FeFET based arrays (Fig. 5 (b)). On applying WAGONN, we observe the inference accuracy increases from 84.85% to 87.53% in 128x128 FeFET array. From Fig. 5 (b), we can observe that WAGONN enables a more graceful decline in accuracy for the 256x256 FeFET array, compared to the sharp decrease observed in baseline designs.

### 5.2 Analysis of WAGONN with PWA/DPWA

Next, we show the effect of combining PWA/DPWA with WAGONN in Fig. 6. We perform this analysis for 128x128 8T SRAM array and fix the number of activated WLs per cycle to 64 in each of the PWA/DPWA cycles.

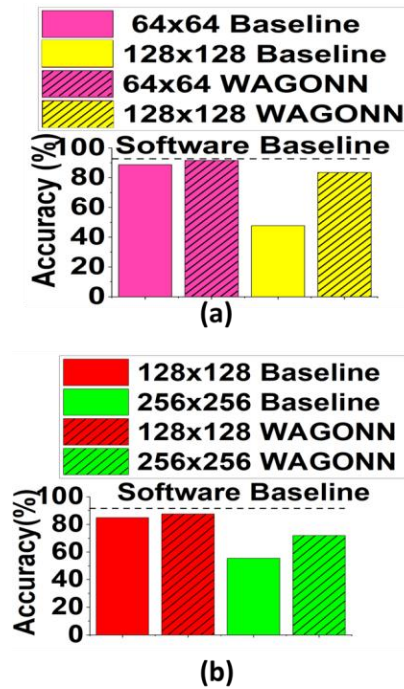

Fig. 6 Accuracy comparison of WAGONN and baseline for (a) SRAM and (b) FeFET for different array sizes

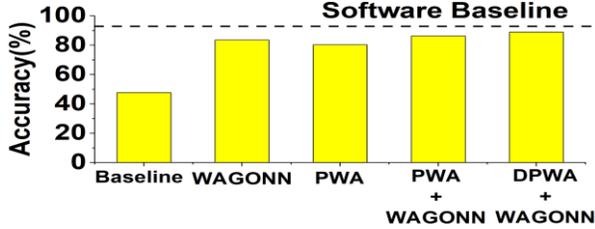

Fig. 7 Comparison of Accuracy using WAGONN with PWA/DPWA on 128x128 8T-SRAM array

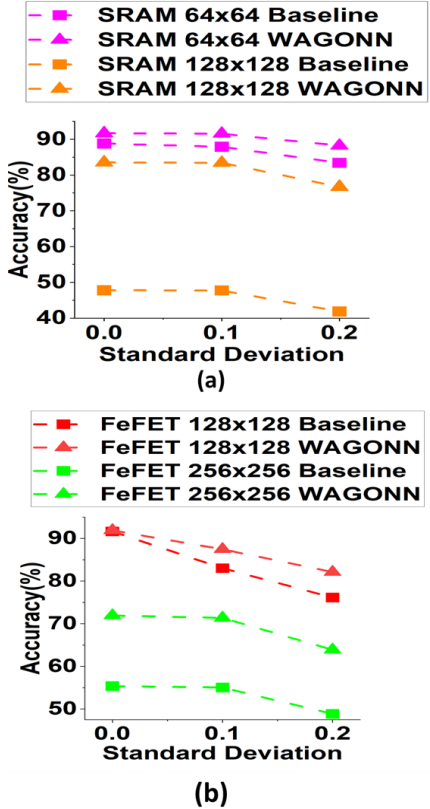

Fig. 8. Effect of variations on baseline design and WAGONN for (a) 8T SRAM and (b) FeFET

Applying WAGONN to a 128x128 SRAM improves accuracy 47.78% to 83.5% while applying just PWA increases accuracy to 80.34% (lower than WAGONN). With PWA + WAGONN, 86.25% is achieved while with the proposed DPWA+ WAGONN, accuracy increases to 88.83%. Higher accuracy in DPWA+WAGONN compared to PWA+WAGONN establishes the synergy between DPWA and WAGONN discussed in Section 4.2.

### 5.3 Analysis of WAGONN with variations

In this section, we analyze the effect of device-to-device variations on the performance of WAGONN. We vary the conductance of each bit-cell by (i) sampling from a Gaussian distribution with a mean of 1 and considering various standard deviations ($s$) and (b) multiplying the bit-cell conductance in the crossbar with these values. This enables us to change the ON as well as the OFF conductance of the bit-cells We use $s = 0$ (no variations), 0.1, and 0.2 to illustrate the trends in inference accuracy for WAGONN and the baseline design. From Fig. 7, we observe that even in the presence of variations, WAGONN consistently provides a higher inference accuracy than the baseline design.

### 5.3 Benchmarking WAGONN against other techniques

In Table II, we benchmark WAGONN against prior non-ideality mitigation solutions. Techniques [6], [7], [8], and [9] are based on non-ideality-aware training, which is computationally expensive. The techniques in [6] and [7] exhibit higher accuracy improvements than WAGONN, that is partly due to much lower wire resistance used (corresponding to older technologies) compared to our work. Furthermore, the benefits in [6]-[7] have been demonstrated on less complex datasets and networks. Compared to [8] and [9], WAGONN achieves comparable/higher improvements despite using a higher wire resistance.

The works in [9] and [11] target the drain-input configuration of the crossbar arrays and employ column rearrangement to position critical columns closer to the drivers. However, the IR-drop effects on the vertical SLs, which influences both $V_{GS}$ and $V_{DS}$ biasing, remain unaddressed. As a result, they exhibit limited accuracy improvements of 7% and 3.5%. On the other hand, WAGONN, improves accuracy by 35.27% (16.35%) for SRAM (FeFET) arrays.

The work in [10] relies exclusively on PWA to improve accuracy but assumes ideal wires with zero resistance. To enhance inference accuracy by 73.4%, approaching software baseline levels, it suggests asserting only one word line (WL) per MVM cycle, which incurs a significant latency penalty. In contrast, WAGONN + DPWA improves accuracy for SRAM (1T-1R) by 41.05% (6.13%) while achieving accuracy very close to the software baseline. WAGONN achieves this in just 2 MVM cycles, as the weight rearrangement within the crossbar—enabled by WAGONN—complements PWA (as discussed in Section 4.2), delivering these accuracy gains with much lower impact on parallelism compared to [10].

## 6 OVERHEAD ANALYSIS

**TABLE II**
Benchmarking WAGONN with other techniques

| Work | Re-train | Resistance per bitcell | MVM cycles | Tech node | Max Array Size | Network | Dataset | Max Accuracy Improvement |
|---|---|---|---|---|---|---|---|---|
| [6] | Yes | 2.5 Ω | 1 | 0T-1R | 784x10 | 2 layer | MNIST | 51.2% |
| [7] | Yes | 2.5 Ω | 1 | 0T-1R | 16x256 | Hopfield | Finger print | 27% |
| [8] | Yes | 10 Ω | 1 | 45 nm 1T-1R | 64x64 | VGG 11 | CF 10/100 | 17% |
| [9] | Yes | 10 Ω | 1 | 0T-1R | 64x64 | VGG 11/16 | CF 10/100 | 7% |
| [11] | No | 2 Ω | 1 | 65 nm 1T-1R | 128x128 | VGG 16 | CF 10 | 3.5% |
| WAGONN (This work) | No | 20 Ω | 1 | 7 nm 8T SRAM | 128x128 | RN-20 | CF 10 | 35.27% |
| WAGONN (This work) | No | 20 Ω | 1 | 7 nm FeFET | 256x256 | RN-20 | CF 10 | 16.35% |
| [10] | No | 0 Ω | 512 | 0T-1M | 512x512 | BCNN | MNIST/SVHN/CF 10 | 73.4% |
| WAGONN + DPWA (This work) | No | 20 Ω | 2 | 7 nm 8T SRAM | 128x128 | RN-20 | CF 10 | 41.05% |
| WAGONN + DPWA (This work) | No | 20 Ω | 2 | 7 nm FeFET | 128x128 | RN-20 | CF 10 | 6.13% |

R – RRAM, M – MTJ, RN - ResNet, CF - CIFAR

## 6.1 Overhead Analysis

We estimate the overheads of WAGONN using SAMBA (Fig. 8) [26], a DNN inference accelerator with a spatial architecture similar to ISAAC [25]. SAMBA features a hierarchical structure comprising Nodes, Tiles, Cores, and MVMUs. The SAMBA accelerator includes multiple tiles connected through an on-chip network that facilitates inter-tile communication. Within a tile, multiple cores share a memory, while each core contains multiple MVMUs, a register file, and a vector function unit. The core handles data loading and storage from/to shared memory within the tile, performs MVM operations to generate partial sums using MVMUs, and executes vectorized arithmetic operations such as partial sum accumulation, normalization, and non-linear activation functions like ReLU. An MVMU includes input/output registers, multiple crossbars, and various peripherals.

The performance of SAMBA is evaluated using a compiler and a cycle-accurate simulator. Compiler maps DNN models to MVMUs and generates the corresponding instructions for cores and tiles. The simulator then runs based on these instructions, measuring overall latency and energy consumption on all calculations and data movement between cores and tiles. Note that while SAMBA suggests multiple data-movement optimizations, we don't utilize them in this work as we focus on evaluating the impact of WAGONN on a generic weight-stationary architecture. While re-mapping weights incurs a one-time cost that can be handled before runtime, re-mapping inputs is dynamic at run-time.

Typically, weight-stationary architectures store inputs in input register and share them across multiple crossbars (Fig. 8). For instance, SAMBA performs input re-mapping within the VFU (outside the MVMU), with the re-mapped inputs then shared among all crossbars within an MVMU. In contrast, WAGONN requires each crossbar to have a uniquely re-mapped input. One option is to use all-to-all muxes but our analysis shows significant latency and routing overheads. Hence, to dynamically re-map inputs with low overheads, we propose the Input Re-mapping Unit (IRU) (Fig. 9), which receives input data from the input register, re-maps it and then provides it to the crossbars. The IRU consists of Look-Up Tables (LUTs) that store the tracking vectors and re-mapped registers that store the re-mapped inputs.

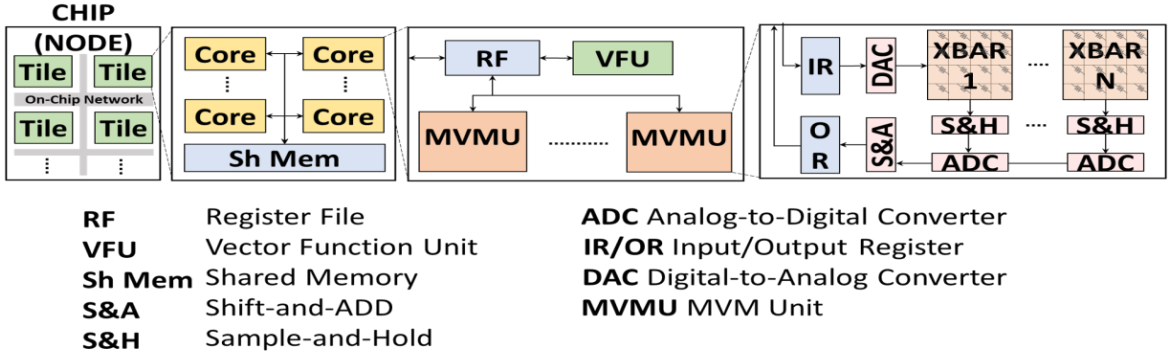

Fig. 9. SAMBA Architecture for DNN accelerators

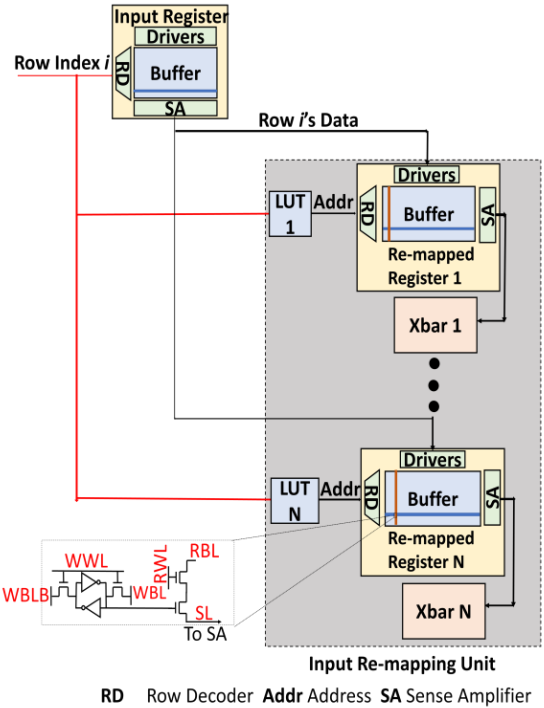

Fig. 10 Architecture of Input Re-mapping Unit

The operation to re-map row $i$ into the re-mapped registers proceeds as follows: First, row $i$'s data is read from the input register. Simultaneously, the LUT is accessed with row index $i$ to determine the destination address of this row (based on the tracking vectors of WAGONN) in the re-mapped register. Then, this destination address is provided to the row decoder and the corresponding data is transmitted to the bit-line drivers of the re-mapped register. Thus, it takes

**Table III**
Area/Energy/Latency Overheads of WAGONN

| Xbar Type | ADCs Per Xbar | Energy Cost (%) | Latency Cost (%) | Area Cost (%) |
|---|---|---|---|---|
| SRAM | 1 | 0.4 | 1.2 | 12.1 |
| SRAM | 16 | 0.6 | 18.8 | 2.7 |
| FeFET | 1 | 0.4 | 1.2 | 15.7 |
| FeFET | 16 | 0.6 | 18.8 | 2.9 |

128 such cycles (of Read and Write) to re-map all inputs of a 128x128 crossbar. Note that each crossbar has a corresponding LUT and re-mapped register in the IRU. Hence, in 128 cycles we can re-map *all* crossbars (which shared the inputs in the baseline design). Although multiple copies of input buffer are required, the additional area consumed by LUTs and re-mapped registers in IRU is not a significant cost (as discussed later), as the ADCs dominate the overall area.

The data in a row of a re-mapped register contains the inputs for the crossbar row. Consequently, input bits of the same significance or place value are aligned along the columns. Therefore, to stream inputs to crossbars, we read the columns of the re-mapped registers. Recall, the write in the remapped registers occurs along the rows. The column readout and row write is naturally feasible in 8T SRAM (Fig. 9), as the read and write paths are separable.

As ADCs consume a dominant share of energy and a substantial share of area of MVM macros [19], the overall overheads of WAGONN can be significantly influenced by the number of ADCs per crossbar. Based on the designs in [25], [27], we analyze the overheads of WAGONN for 1 ADC and 16 ADCs per crossbar.

We evaluate these overheads for SAMBA for two designs: (i) 128x128 and (ii) 128x128 FeFETs with 8T SRAM-based LUT and remapped registers. We do not design LUTs and re-mapped registers with FeFETs due to high write time/energy compared to SRAM and limited endurance [4]. From Table III, it is evident that WAGONN incurs minimal energy overhead of at most 0.6%. This minimal overhead is primarily due to the dominance of ADCs in the overall energy consumption, which diminishes the relative energy cost of the IRU required in the WAGONN design.

When 1 ADC is used per crossbar array, WAGONN incurs at most 1.2% latency cost. This minimal overhead occurs from sharing an ADC among multiple columns in a crossbar array which increases the overall latency, even for the baseline design, thus reducing the relative impact of additional circuitry of WAGONN. However, when 16 ADCs are used per crossbar array, the baseline design latency significantly decreases, resulting in WAGONN's latency overhead increasing to 18.8%.

We also observe that the area overhead of WAGONN is up to 15.7% with one ADC per crossbar and 2.9% with 16 ADCs per crossbar. As the number of ADCs per crossbar increases, they occupy a larger proportion of the accelerator's area, thereby reducing the relative area share of the additional circuitry required by WAGONN. For 256 x 256 FeFET, WAGONN overheads will be even lower as ADC conversion time and energy increases non-linearly with precision.

## 6 CONCLUSION

We propose WAGONN to mitigate interconnect-induced non-idealities and enhance DNN inference accuracy by re-mapping weights in crossbars. By sorting rows based on their row-sum, WAGONN places most low-resistance state bit-cells (of 8T SRAMs/FeFETs) closest to the ADC, reducing wire resistance-induced source degeneration. At 7nm, we show that WAGONN improves the inference accuracy of 128x128 SRAM by 35.72% compared to the baseline. We also show accuracy improvements for FeFETs. Further, we synergistically utilize WAGONN with DPWA, resulting in improved accuracy compared to using these techniques individually. WAGONN incurs minimal hardware overhead, with less than 1% increase in energy consumption. Additionally, the latency overhead of WAGONN is ~ 1%, while the area overhead is ~ 16% when 1 ADC is used per crossbar array.

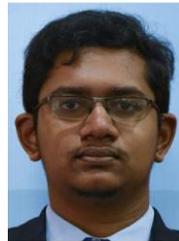
**Jeffry Victor** is a PhD candidate at the Department of Electrical and Computer Engineering, Purdue University, under the supervision of Sumeet Gupta. His research focuses on the cross-layer design of compute-enabled memories for emerging neural workloads. He has a bachelor's degree in technology from Birla Institute of Technology and Science Pilani, India.

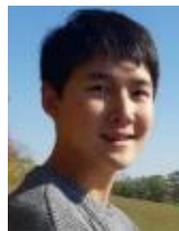
**Dong Eun Kim** received his BS degree in electrical and electronic engineering from Yonsei University, Korea, in 2018. He is a PhD candidate in electrical and computer engineering with Purdue University, under the supervision of professor Kaushik Roy. His research interests include hardware-software co-design for neural network accelerators. During his PhD, he interned with IBM in 2021.

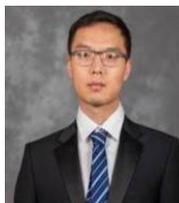
**Chunguang Wang** is a PhD candidate at the Department of Electrical and Computer Engineering, Purdue University, under the supervision of Sumeet Gupta. His research focuses on the device-circuit interactions of emerging NVMs. He has a bachelor's and master's degree from Peking University, China. During his PhD, he interned with Micron in 2024.

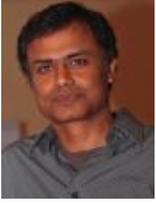
**Kaushik Roy** (Fellow, IEEE) is the Edward G. Tiedemann, Jr., Distinguished Professor of Electrical and Computer Engineering at Purdue University. He received his BTech from the Indian Institute of Technology, Kharagpur, PhD from the University of Illinois at Urbana-Champaign in 1990. He joined the Semiconductor Process and Design Center of Texas Instruments, Dallas, where he worked for three years on FPGA architecture development and low-power circuit design. His current research focuses on cognitive algorithms, circuits and architecture for energy-efficient neuromorphic computing/ machine learning, and neuromimetic devices. Kaushik has supervised more than 100 PhD dissertations and his students are well placed in universities and industry. He is the co-author of two books on Low Power CMOS VLSI Design (John Wiley & McGraw Hill). Throughout his career, Dr. Roy received several awards including the National Science Foundation Career Development Award in 1995, IBM faculty partnership award, ATT/Lucent Foundation award, 2005 SRC Technical Excellence Award, SRC Inventors Award, Purdue College of Engineering Research Excellence Award, Outstanding Mentor Award in 2021, Humboldt Research Award in 2010, 2010 IEEE Circuits and Systems Society Technical Achievement Award (Charles Desoer Award), IEEE TCVLSI Distinguished Research Award in 2021, Distinguished Alumnus Award from Indian Institute of Technology (IIT), Kharagpur, Fulbright-Nehru Distinguished Chair, DoD Vannevar Bush Faculty Fellow (2014-2019), SRC Aristotle Award in 2015, Purdue Arden L. Bement Jr. Award in 2020, SRC Innovation Award in 2022, and an honorary doctorate from Aarhus University in 2023.

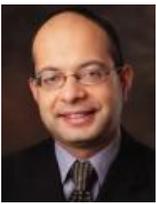
**Sumeet Kumar Gupta** (Senior Member, IEEE) received the B.Tech. degree in electrical engineering from the Indian Institute of Technology Delhi, New Delhi, India, in 2006, and the M.S. and Ph.D. degrees in electrical and computer engineering from Purdue University, West Lafayette, IN, USA, in 2008 and 2012, respectively. He is currently an Elmore Associate Professor of Electrical and Computer Engineering with Purdue University. Prior to this, he was an Assistant professor of Electric Engineering with The Pennsylvania State University, State College, PA, USA, from 2014 to 2017, and a Senior Engineer with Qualcomm Inc., San Diego, CA, USA, from 2012 to 2014. He has also worked as an Intern with National Semiconductor, Santa Clara, CA, USA, in 2005; Advanced Micro Devices Inc., Santa Clara, in 2007; and Intel Corporation, Hillsboro, OR, USA, in 2010. He has published over 100 articles in refereed journals and conferences. His research interests include low-power variation aware VLSI circuit design, neuromorphic computing, in-memory computing, nanoelectronics and spintronics, device-circuit co-design, and nanoscale device modeling and simulations. Dr. Gupta was the recipient of the DARPA Young Faculty Award in 2016, the Early Career Professorships by Purdue and Penn State in 2021 and 2014, respectively, and the 6th TSMC Outstanding Student Research Bronze Award in 2012. He has also received the Magoon Award and the Outstanding Teaching Assistant Award from Purdue University in 2007 and Intel Ph.D. Fellowship in 2009. He is a Senior Member of EDS.